\documentclass[twocolumn,amsmath,amssymb,pra,a4paper]{revtex4}

\usepackage{graphicx}
\usepackage{float}
\usepackage{epsfig}
\usepackage{subfigure}
\usepackage{pstricks}
\usepackage{color}

\newcommand{\Ga}{\Gamma}

\newcommand{\Ev}{\mathcal{E}}

\newcommand{\vq}{{\vec q}}
\newcommand{\vp}{{\vec p}}

\newcommand{\ket}[1]{|{#1}\rangle}

\newcommand{\bracket}[2]{\langle{#1}|{#2}\rangle}

\newcommand{\half}{\frac{1}{2}}

\newcommand{\re}{\mathop{\text{Re}}}

\date{\today}
\begin{document}
\title{Dynamical phase interferometry of cold atoms in optical lattices}
\author{Uri London and Omri Gat}
\affiliation{Racah Institute  of  Physics, Hebrew University
of Jerusalem, Jerusalem 91904, Israel}
\begin{abstract}
We study the propagation of cold-atom wave packets in an interferometer with a Mach-Zehnder topology based on the dynamical phase of Bloch oscillation in a weakly forced optical lattice with a narrow potential barrier that functions as a cold-atom wave packet splitter. We calculate analytically the atomic wave function, and show that the expected number of atoms in the two outputs of the interferometer oscillates rapidly as a function of the angle between the potential barrier and the forcing direction with period proportional to  the external potential difference across a lattice spacing divided by the lattice band energy scale. The interferometer can be used as a high precision force probe whose principle of operation is different from current interferometers based on the overall position of Bloch oscillating wave packets.
\end{abstract}

\maketitle

\section{Introduction}
The dynamics of a quantum particle in a periodic environment is in many ways similar to that of a free particle, but when subject to a uniform force it does not accelerate indefinitely but performs periodic motion---Bloch oscillations. This effect is often explained using interference of Wannier-Stark resonances, but can also be understood in terms of classical dynamics generated by an effective Hamiltonian, where the kinetic term is replaced by the band dispersion---Peierls substitution \cite{peierls,kohn,blount,zak68,niudyn,pst}. The band dispersion is periodic so that the direction of the velocity, that is the gradient of the Hamiltonian with respect to momentum, oscillates, and the classical trajectories generated by the effective Hamiltonian are periodic.

The high controllability and weak coupling to the environment of cold atom systems makes them ideal for observation of Bloch oscillation \cite{bendahan,niu,kasevich,morsch,rmp,korsch}. One of the experimental applications of cold-atom Bloch oscillations has been force measurements, in particular of the acceleration of gravity \cite{clade,roati,fattori,ferrari,gustavsson,poli}, that has recently achieved $10^{-7}$ accuracy. This application is based on a precise knowledge of the periodic potential and a large number of oscillation that enable an accurate measurement of the period. The experiments are carried out with very weakly interacting Bose atoms to minimize dephasing.

\begin{figure}[htb]
{\epsfig{file=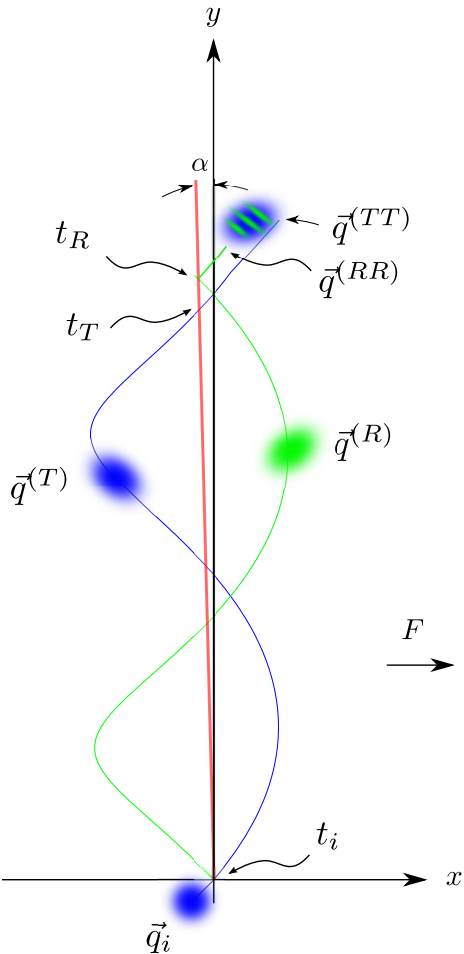,width=0.7\linewidth}}
\caption{\label{fig:int}(Color online) A schematic diagram of the cold atom interferometer, consisting of  a square optical lattice whose axes are shown as the $x$ and $y$ axes, an external force $\vec F$ in the $x$ direction, and a tunnel barrier (red) tilted by a small angle $\alpha$ with respect to the $y$ axis. The initial wave packet indicated at the bottom of the figure (blue) at phase space point $(\vq_i,\vp_i)$ splits at time $t_i$ by scattering into a transmitted and reflected sub-wave packets, whose classical reference orbits $(\vq_i^{(T)},\vp_i^{(T)})$ and $(\vq_i^{(T)},\vp_i^{(T)})$ (respectively) are indicated in blue and green (respectively), that scatter once more from the tunnel barrier at times $t_T$ and $t_R$ (respectively) and recombine. The interference of the transmitted-transmitted (TT) and reflected-reflected (RR) wave packets is sketched near the top of the figure. Some examples of the expected fraction of the atoms in the TT-RR output are shown in Figs.\ \ref{fig:output-1d} and \ref{fig:output-2d}.}
\end{figure}

A cold-atom wave packet undergoing Bloch oscillations in an optical lattice with an external force accumulates a dynamical phase during its propagation. The dynamical phase accumulates much faster than the oscillations phase, so that a {single} Bloch oscillation yields a high-precision measurement of the external force.  However, the measurement of a phase requires an interferometric setup. For this purpose we propose to use a narrow one dimensional potential barrier as a cold-atom wave packet splitter to form an interferometer with a Mach-Zehnder topology, shown schematically in Fig.\ \ref{fig:int}. The auxiliary mirrors of optical Mach-Zehnder interferometers are not required here, since the trajectories in the two arms of the cold atom interferometer are curved, and naturally recombine.

A tunnel barrier for cold atoms can be implemented by a narrow blue-detuned laser beam. Coherent tunneling of Bose-Einstein condensates through a barrier significantly narrower than the spatial extent of the condensate has been used in experiments to study condensate dynamics in a double-well potential \cite{oberthaler,immanuel,jeff}. Here we propose to use the same technique in an optical lattice instead of a trap. The width of the barrier, equal to several lattice spacings, fits the requirements of the interferometer.

The proposed setup is therefore as follows: a cold-atom wave packet is loaded adiabatically into an optical lattice with lattice spacing much shorter than the (spatial) wave packet width, so that only a single energy band is occupied. We assume that the band is non-degenerate and does not cross other bands. The atoms are also subject to a weak external force along a lattice direction. The wave packet is accelerated by an external force and impinges on a narrow potential barrier that is nearly perpendicular to the force direction; it then splits, the wave packet fragments propagate in the two arms, and recombine in a second collision with the potential barrier. The interferometer structure is summarized in Fig.\ \ref{fig:int}. High sensitivity is derived from the fast oscillations of the interferometer outputs as a function of the tilt angle of the tunnel barrier.

The interferometer system relies on three small parameters. The principal parameter is $\varepsilon$, the ratio of the potential energy difference across a lattice unit divided by the energy band energy scale. The inequality $\varepsilon\ll1$ guarantees the validity of the semiclassical approximation and provides the necessary scale separation between propagation and scattering in the operation of the interferometer. It is also essential for the interferometric sensitivity of the phase, since the accumulated phase difference in the two arms is proportional to $\varepsilon^{-1}$. The second small parameter is $\delta$, the ratio of the lattice spacing and the spatial uncertainty of the cold-atom wave packet. $\delta\ll1$ is necessary for the localization of the wave packet in a single band, while momentum localization requires that $\varepsilon\ll\delta$. The third small parameter is the lattice spacing divided by the width $w$ of the tunnel barrier. $w$ has to be much shorter than the wave packet width to prevent wave packet distortion during the scattering and much longer than $a$ to prevent band transitions. The setup conditions are therefore
\begin{equation}\label{eq:ie}
\varepsilon\ll\delta\ll\frac a w\ll1
\end{equation}

Good visibility of the interference between the two wave packets in the outputs of the interferometer depends on phase-space overlap. Time-reversal symmetry guarantees that the transmitted and reflected wave packets recombine at the beam splitter with full overlap after a complete Bloch oscillation if the barrier is perpendicular to the force direction, \emph{regardless} of the band dispersion; it follows by continuity that there is significant overlap when the tilt angle $\alpha$ of the beam splitter with respect to this direction is small enough. The population in the outputs exhibits fast oscillations as a function of $\alpha$ with a period of order $\varepsilon$, shown in Eq.\ (\ref{eq:ttfcrr}) and Figs.\ \ref{fig:output-1d} and \ref{fig:output-2d}. The range of $\alpha$ with high visibility is of order $\varepsilon^{\frac12}$ for the optimal choice $\delta\sim\varepsilon^{\frac12}$, so that the number of visible oscillation is large, $\sim \varepsilon^{-\frac12}$. In the initial, constant visibility range of $\alpha$, the oscillations are purely sinusoidal with a period determined by the initial momentum and the band dispersion, as shown in Eq.\ (\ref{eq:oalpha-ex}).

The wave packet motion in the arms of the interferometer is planar, undergoing Bloch oscillations in one direction and free motion in the perpendicular direction. A 1D optical lattice is therefore required, with lattice axis along the direction of the external force. Nevertheless, the interferometer operates equally well in a 2D lattice, and such operation offers additional probes, discussed below. Since the 1D-lattice case is obtained as special case of the 2D one, we carry out the analysis for a 2D lattice, and present the 1D special case for the key results.

The recombination of the wave packet occurs after a single Bloch oscillation. However, the interferometer geometry implies that the wave packets propagating in the interferometer arms cross the barrier once also during the Bloch period. The wave packets in general do not recombine during this intermediate crossing, so it is useless for the purpose of interferometry. A practical method of avoiding these spurious collisions and subsequent degradation of the interference pattern is to make the barrier potential time-dependent. The barrier potential needs to be maintained only during the initial and final collisions, and can be turned off  during the intermediate crossing of the barrier, since these are well-separated in time. We will assume that this plan is carried out in our calculations and, for the sake of completeness, point out the modifications that arise if the barrier is time-independent.

Bloch oscillations are a wave phenomenon that is not particular to quantum mechanics. In addition to cold atoms systems they have been experimentally observed and studied in guided light systems \cite{hagai,lederer}. These systems are typically paraxial, and since the paraxial wave equation is identical to the Schr\"odinger equation with the propagation direction playing the role of the time coordinate, the propagation of a quantum wave packet in a periodic potential and of a beam in wave guide array are described by the same analysis. This result also holds for the present study; for concreteness we express all our results in cold atom terms, as it is the more natural application. 

In addition to the spatial interferometry considered here, waves propagating in a periodic medium can split and interfere in the band degree of freedom \cite{bz,longhi-short,longhi-long,longhi-exp}. Instead of a tunnel barrier, this requires a coupling between bands, possibly by Landau-Zener tunneling. This type of interference does not directly measure the external force unless the periodic structure is controlled. 

The interferometer operation is based on two building blocks: Semiclassical wave packet propagation \cite{lj} in the arms of the interferometer, that is presented in Sec. \ref{sec:wp}, and quantum normal form analysis \cite{wws} of the scattering of the wave packets off the tunnel barrier, that is the subject of Sec.\ \ref{sec:scat}. The results of the wave packet propagation analysis and wave packet scattering are combined in Sec.\ \ref{sec:int} to yield expressions for the interferometer output as a function of the tilt angle. In Sec.\ \ref{sec:cc} we discuss applications of the interferometer  force measurement and  as a probe of the optical lattice structure. It is shown that the interferometer is experimentally realizable for a standard setup with gravity as the external force

\section{Wave packet Bloch oscillations}\label{sec:wp}
The propagation of the wave packets in the arms of the interferometer is governed by the combination of the optical lattice potential, and the external force $\vec F$ directed along a lattice axis, that we choose as the $x$ axis, so that the single-atom Hamiltonian is
\begin{equation}
\frac{\vec{\mathsf{p}}^2}{2m}+V_\text{per}(\vec q)-\vec q\cdot\vec F
\end{equation}
where $\vec q$ and $\vec{\mathsf{p}}$ are the atom's position and momentum, respectively, and $V_\text{per}$ is the lattice potential.
We assume that the inter-atom interactions are negligible, and that the many-atom state is a product of identical one-atom states

The optical lattice is either one- or two-dimensional. In either case, we suppose that the wave packet is fully supported in one of the energy bands of the (unforced) lattice potential whose energy dispersion is $\Ev(p_x,p_y)$, where $\vp$ is the quasi-momentum, and measure energy in units of $U$, the energy band width. If the optical lattice is one-dimensional then $\Ev(p_x,p_y)=\Ev_1(p_x)+\frac{1}{2m}p_y^2$, where $\Ev_1$ is the one-dimensional dispersion, and $p_y$ is the $y$ component of the ordinary momentum. We assume that the energy band is nondegenerate and separated from the other bands by a gap of order $U$. We also choose $\frac h a$, where $a$ is the lattice spacing, as the unit of momentum, so that $\Ev$ is periodic with period one in $p_x$. The external force is weak, in the sense that $\varepsilon\equiv\frac{aF}{U}\ll1$. In the following we will write simply momentum instead of quasi-momentum, when there is no risk of confusion. 

The initial single-atom state $\ket{i}$ is a wave packet with mean position $\vq_i$ and momentum $\vp_i$ and momentum uncertainty $\delta\ll1$. It follows that the position uncertainty is much larger than $a$. In explicit calculations we let $\ket{i}$ be a Gaussian wave packet with wavefunction $\bracket{\vp}{i}=c\exp(-\frac{(\vp-\vp_i)^2}{4\delta^2}-\frac{i}{\hbar}\vq_i\cdot\vp)$.

Under these conditions, an effective classical dynamics can be applied as a systematic approximation for the propagation of the wave packet. An effective Hamiltonian $H_\text{eff}$ determines the in-band dynamics, derived from a semiclassical expansion in $\varepsilon$, whose leading term, Peierls substitution \cite{peierls}, is obtained by replacing the kinetic term in the Hamiltonian with the band dispersion, so that
\begin{equation}\label{eq:heff}
H_\text{eff}=\Ev(\vp)-\vq\cdot\vec F
\end{equation}
and letting $\vq$ and $\vp$ be canonically conjugate $[q_n,p_m]=i\hbar\delta_{nm}$. $\vq$ is a coarse-grained position observables with an uncertainty much larger than $a$. We assume that the optical lattice is space-reflection invariant in $x$ and $y$, so that the sub-leading term in $H_\text{eff}$ is $O(\varepsilon^2)$ and therefore negligible.

The semiclassical propagation of the wave packet generated by $H_\text{eff}$ consists of three elements, a phase-space drift along the classical trajectory of the center of the wave packet $(\vq_t,\vp_t)$, a deformation (squeezing) of the wave packet by the linearized flow at the wave packet center, and an overall phase accumulation proportional to the classical action of the wave packet center trajectory \cite{heller76,lj}. Each part is realized as a unitary operator, so that the wave packet at time $t_f$ can be represented by
\begin{equation}\label{eq:scwp}
 \ket{f}=e^{i\gamma_{fi}}T(\vq_f,\vp_f)M(S_{fi})T^\dagger(\vq_i,\vp_i)\ket{i}
\end{equation}
where 
$$\gamma=-{\frac{1}{\hbar}}\int_i^f(\textstyle\frac12(\vp\cdot d\vq-\vq\cdot d\vp)-H_\text{eff}dt)$$
is the dynamical phase, 
$$T(\vq_0,\vp_0)= \textstyle\exp(\pm\frac{i}{\hbar}(\vq\cdot\vp_0-\vp\cdot\vq_0))$$
is the Heisenberg phase-space shift operator, and $M(S)$ is the squeezing (metaplectic) operator derived from the symplectic phase-space deformation matrix
$$S_{fi}=\left(\begin{matrix}S_{qq}&S_{qp}\\S_{pq}&S_{pp}\end{matrix}\right)$$ 
with block components $S_{qq}=\partial q_{f,n}/\partial q_{i,m}$, $S_{pq}=\partial p_{f,n}/\partial q_{i,m}$, $S_{qp}=\partial q_{f,n}/\partial p_{i,m}$, and $S_{pp}=\partial p_{f,n}/\partial p_{i,m}$. 

The problem is therefore reduced to the calculation of the classical trajectories of 
$H_\text{eff}$. The effective Hamiltonian is integrable, and its flow can be calculated explicitly. The Hamilton equations are
\begin{align}
 \partial_t\vq&=\nabla\Ev(\vp)\\
 \partial_t\vp&=\vec F
\end{align}
The momentum equation implies that $\vp_t=\vp_i+\vec Ft$; the $y$ position equation $q_{t,y}=q_{i,y}+\int_{t_i}^tdt' v_y(t')$, where $v_y(t)=\partial_y\Ev(\vp_t)$, and conservation of energy implies that $q_{t,x}=q_{i,x}+\frac1F(\Ev(\vp_t)-\Ev(\vp_i))$. The periodicity of $\Ev$ implies that the wave packet performs Bloch oscillations in the $x$ direction with period $\frac 1F$, while moving freely in the $y$ direction.
 
It follows that the $S_{qq}$ and $S_{pp}$ blocks of the deformation matrix are unit matrices, the $S_{pq}$ block is the zero matrix, and the elements of the $S_{qp}$ block are
\begin{align}
\frac{\partial q_{f,x}}{\partial p_{i,x}}&=\frac 1F(v_x(\vp_f)-v_x(\vp_i))\\
\frac{\partial q_{f,y}}{\partial p_{i,x}}&=\frac{\partial q_{f,x}}{\partial p_{i,y}}=\frac 1F(v_y(\vp_f)-v_y(\vp_i))\\
\frac{\partial q_{f,y}}{\partial p_{i,y}}&=\frac 1F\int_i^fdp_x \partial_y v_y(p_x,p_{i,y})
\end{align}
The expression for the dynamical phase can be simplified to
\begin{equation}
\gamma_{fi}=\frac1{2\hbar}(\vp_f\cdot\vq_f-\vp_i\cdot\vq_i)-\frac{U}{\hbar F}\int_i^fdp_x\Ev(p_x,p_{i,y})
\end{equation}

In our choice of units the coefficient $\frac{U}{\hbar F}=\frac{2\pi}{\varepsilon}$, so that the classical limit, where $\hbar\to0$ keeping all classical quantities fixed, is equivalent here to the limit $\varepsilon\to0$, and $\varepsilon$ is the effective Planck constant. In this limit the dynamical phase diverges, giving rise to the high interferometric sensitivity, as the small wavelength does in classical optical interferometry.

\section{Wave packet splitting off a tunnel barrier}\label{sec:scat}
The atom wave packet is split by tunneling through a one-dimensional potential barrier or a potential trench, that is, a modulation of the optical potential of the form $u_b(\vq\cdot\hat n)$, where $\hat n$ is the unit normal to the beam splitter, and $u_b$ is localized on a scale $w$ of magnitude $a\ll w\ll a/\delta$.

For the purpose of analyzing the scattering of the wave packet by the beam splitter, we ignore temporarily the external force, and consider a standard scattering problem of an incoming unit amplitude plane wave with quasi-momentum $\vp$ in an energy band with dispersion $\Ev$, that splits into a transmitted wave with quasi-momentum $\vp$ with amplitude $t$ and a reflected wave with quasi-momentum $\vp^{(R)}$ with amplitude $r$. The inequality $a\ll w$ guarantees that the scattered wave remains in the same energy band. Lattice translation symmetry implies that tangential component of quasi-momentum is conserved, $p_t=p_t^{(R)}$, and energy conservation implies that $\Ev(\vp)=\Ev(\vp^{(R)})$; we assume that the band dispersion is such that there is a single solution $\vp^{(R)}$, other than $\vp$, to these two conditions.

The initial quasi-momentum and wave packet splitter orientation define a classical band of allowed energies between $\Ev_-=\min_{p_n}\Ev(\vp+p_n\hat n)$ and $\Ev_+=\max_{p_n}\Ev(\vp+p_n\hat n)$. The condition $a\ll w$ also implies that unless $\Ev(\vp)-\Ev_-$ is close to $\max u_b$ or $\Ev(\vp)-\Ev_+$ is close to $\min u_b$ either $|t|\ll1$ or $|r|\ll1$. For  concreteness we choose to study scattering from a potential barrier and let  $|\Ev(\vp)-\Ev_--\max u_b|\ll1$ so that both $t$ and $r$ are appreciable to obtain good interferometric visibility. We also assume that the initial momentum is such that $\Ev(\vp_i)$ is far from the edges $\Ev_\pm$ of the interval of allowed energies.

The typical barrier maximum and dispersion minimum are quadratic, and under the conditions laid out above the scattering amplitudes are determined by local behavior of $\Ev$ and $u_b$ near their extrema \cite{wws}. Denoting by $\vp_m$ and $q_m$ the abscissas of the extrema, we define the classical action
\begin{equation}
I=\frac{\Ev(\vp)-\Ev(\vp_m)-u_b(q_m)}{\sqrt{-\partial_n^2\Ev(\vp_m)\partial_q^2u(q_m)}}
\end{equation}
and the standard theory of scattering from a parabolic barrier gives \cite{mount}
\begin{align}
t(\vp) =& \frac{e^{i\frac{I}{\hbar}(\log\frac{I}{\hbar}-1)}}{\sqrt{2\pi}}\Ga\Big(\half-i\frac{I}{\hbar}\Big)e^{\frac{\pi}{2}\frac{I}{\hbar}}\label{eq:tp}\\
r(\vp) =& -i\frac{e^{i\frac{I}{\hbar}(\log\frac{I}{\hbar}-1)}}{\sqrt{2\pi}}\Ga\Big(\half-i\frac{I}{\hbar}\Big)e^{-\frac{\pi}{2}\frac{I}{\hbar}}
\label{eq:rp}\end{align}

A scattering wave packet is a superposition of stationary scattering states $\ket{i}=\int dp \psi(\vp)\ket{\vp}$. After the center of the incoming wave packet collides with the barrier it splits into a transmitted wave packet $\ket{i}^T=\int d \vp t(\vp)\psi(\vp)\ket{\vp}$ and a reflected wave packet $\ket{i}^R=\int dp r(p)\psi(p)\sqrt{|\det J|}\ket{\vp^{(R)}}$, where $J=\frac{\partial\vp}{\partial\vp}^{\scriptscriptstyle(R)}$ is the Jacobian matrix of the reflection transformation. Since the incoming wave packet is concentrated near $\vp_i$, the transmitted wave packet is approximately 
\begin{equation}\label{eq:iT}
\ket{i}^T=\int dp t(\vp_i)\psi(\vp)\ket{\vp}=t(\vp_i)\ket{i}
\end{equation}
but the dependence of the scattering amplitudes on $\vp$ causes a position-space shift and distortion of the wave packet. Nevertheless, the shift and distortion are weak if the $t$ and $r$ do not change appreciably over the range of momenta that support the wave packet, or equivalently if the barrier width $w$ is much smaller than the position uncertainty $\frac{a}{\delta}$ \cite{segev}, as we assume. In this case the position and momentum shifts incurred by the scattering are of $O(w)$ and $O(\frac{w}{a}\delta^2)$ (respectively), negligible in comparison with the respective uncertainties; the relative deformation is even smaller, of $O\bigl((\frac{w\delta}{a})^2\bigr)$.

By the same reasoning the reflected wave packet can be written as $\ket{i}^R=r(p_i)\int d\vp \psi(\vp)\sqrt{|\det J|}\ket{\vp^{(R)}}$. Since the momentum integration is localized, $J$ can be evaluated at $\vp_i$, and the reflected momentum is approximately $\vp^{(R)}=\vp_i^{(R)}+J_i(\vp-\vp_i)$, where $J_i=J(\vp_i)$. The reflected wave packet is therefore
\begin{equation}\label{eq:iR}
\ket{i}^R=r(p_i)T(\vq_i)T(\vp_i^{(R)})M\bigl(\begin{smallmatrix}(J_i^t)^{-1}&0\\0&J_i\end{smallmatrix}\bigr)T^\dagger(\vp_i)T^\dagger(\vq_i)\ket{i}
\end{equation}

Finally we reconsider the effect of external force.
In this case the quasi-momentum is no longer a good quantum number, so we label the stationary states by the value of $\vp$ at the barrier. These states are propagating in a large interval of $O(\frac{a}{\varepsilon})$ size around the barrier and can therefore be used as a basis for a scattering theory for the wave packets considered here, with position uncertainty of $O(\frac{a}{\delta})\ll\frac{a}{\varepsilon}$. The preceding arguments and Eqs.\ (\ref{eq:tp}--\ref{eq:iR}) are valid also with the weak uniform external force up to small corrections. 

\section{Angle interferometry}\label{sec:int}
The wave packet interferometry takes place in three steps. An initial wave packet $\ket{i}$ of mean position $\vq_i$ and momentum $\vp_i$ is first split by scattering from the wave packet splitter into a transmitted wave packet $\ket{i}^T$ and a reflected wave packet $\ket{i}^R$ with mean position $\vq_i$ and momentum $\vp_i^{(R)}$. The transmitted wave packet then undergoes Bloch oscillation in the $x$ direction and propagates freely in the $y$ direction before impinging on the beam splitter again at time $t_T$ and position $\vq_c^{T}$ with mean momentum $\vp_c^{(T)}$. The wave packet $\ket{c}^{T}$ then splits again into a transmitted-transmitted (TT) wavepacket $\ket{c}^{TT}$ at $\vq_c^{(TT)}$ and $\vp_c^{(TT)}$ and a transmitted-reflected (TR) wave packet $\ket{c}^{TR}$ at $\vq_c^{(TR)}$ and $\vp_c^{(TR)}$. The states and variables related to the propagation and splitting of the reflected wave packet are similarly defined, see Fig.\ \ref{fig:int}.

The system has a Mach-Zehnder geometry with two outputs, one interfering 
TT with RR wave packets, and the other interfering TR with RT wave packets. The interferometer is sensitive to the difference between the dynamical phases accumulated by the transmitted and reflected wavepackets during their propagation. This phase difference changes fast as a function of the tilt angle $\alpha$ between the beam splitter and the $y$ axis.

\subsection{Wave packet interferometry}
We now calculate the interference pattern, concentrating from this point on the TT-RR output. The interference takes place after both the transmitted and the reflected wave packets have collided again with the beam splitter. When the tilt angle $\alpha$ is positive $t_R>t_T$, so that $t_f=t_R$. The final state of the TT wave packet is therefore
\begin{align}
\ket{f}^{TT}&=t(\vp_c^{(T)})t(\vp_i)e^{i\gamma_{fi}^{(TT)}}T(\vq_f^{(TT)},\vp_f^{(TT)})\nonumber\\
&\times M(S_{fi}^{(TT)})T^\dagger(\vq_i,\vp_i)\ket{i}\label{eq:ftt}
\end{align}
with momentum-space wave function
\begin{align}\label{eq:pftt}
\bracket{\vp}{f}^{TT}&=t(\vp_c^{(T)})t(\vp_i) e^{-\frac{2\pi i}{\varepsilon}\int_{\vp_{i,x}}^{\vp_{f,x}^{(TT)}}\Ev(p_x,p_{i,y})dp_x}\nonumber\\
&\times\frac{1}{\sqrt{2\pi\delta^2}}e^{\frac{i}{\hbar}(\vq_f^{(TT)}\cdot\vp_f^{(TT)}-\vq_i\cdot\vp_i)}e^{-\frac{i}{\hbar}\vq_f^{(TT)}\cdot\vp}\nonumber\\&\times e^{-\frac{1}{{4\delta^2}}(\vp-\vp_f^{(TT)})
\cdot C^{(TT)}\cdot(\vp-\vp_f^{(TT)})}
\end{align}
for $C^{(TT)}=1+i\frac{2\delta^2}{\hbar}S_{f,pq}^{(TT)}$,
and the final state of the RR wave packet is
\begin{align}\label{eq:crr}
\ket{f}^{RR}&=\ket{c}^{RR}=r(\vp_c^{(R)})r(\vp_i)e^{i\gamma_{ci}^{(R)}} T(\vq_c^{(R)})\nonumber\\& \times T(\vp_c^{(RR)})M\bigl(\begin{smallmatrix}((J_c^{(R)})^t)^{-1}&0\\0&J_c^{(R)}\end{smallmatrix}\bigr)T^\dagger(\vp_c^{(R)})T^\dagger(\vq_c^{(R)})\nonumber\\& \times
T(\vq_c^{(R)},\vp_c^{(R)})
M(S_{ci}^{(R)})T^\dagger(\vq_i,\vp_i^{(R)})\nonumber\\& \times
T(\vq_i)T(\vp_i^{(R)})M\bigl(\begin{smallmatrix}(J_i^t)^{-1}&0\\0&J_i\end{smallmatrix}\bigr)T^\dagger(\vp_i)T^\dagger(\vq_i)\ket{i}
\end{align}
with wave function
\begin{align}\label{eq:pcrr}
\bracket{\vp}{c}^{RR}&=r(\vp_c^{(R)})r(\vp_i) e^{-\frac{2\pi i}{\varepsilon}\int_{\vp_{i,x}^{(R)}}^{\vp_{c,x}^{(R)}}\Ev(p_x,p_{i,y})dp_x}\nonumber\\
&\textstyle\times\frac{e^{\frac{i}{\hbar}(\vq_c^{(R)}\cdot\vp_c^{(R)}-\vq_i\cdot\vp_i^{(R)})}}{\sqrt{2\pi\delta^2\det(J_i)\det(J_c^{(R)})}}e^{-\frac{i}{\hbar}\vq_c^{(R)}\cdot\vp}\nonumber\\&\times e^{-\frac{1}{{4\delta^2}}(\vp-\vp_c^{(RR)})
\cdot C^{(RR)}\cdot(\vp-\vp_c^{(RR)})}
\end{align}
for
\begin{displaymath} \textstyle
C^{(RR)}=((J_c^{(R)})^t)^{-1}\Bigl((J_i^t)^{-1}J_i^{-1}+i\frac{2\delta^2}{\hbar}S_{c,pq}^{(R)}\Bigr)(J_c^{(R)})^{-1}.
\end{displaymath}
Conservation of energy and transverse momentum imply that in Cartesian coordinates
\begin{align}
J_i&=\begin{pmatrix}v_{i,x}^{(R)}&v_{i,y}^{(R)}\\-\sin\alpha&\cos\alpha\end{pmatrix}^{-1}
\begin{pmatrix}v_{i,x}&v_{i,y}\\-\sin\alpha&\cos\alpha\end{pmatrix}\\
J_c^{(R)}&=\begin{pmatrix}v_{c,x}^{(RR)}&v_{c,y}^{(RR)}\\-\sin\alpha&\cos\alpha\end{pmatrix}^{-1}
\begin{pmatrix}v_{c,x}^{(R)}&v_{c,y}^{(R)}\\-\sin\alpha&\cos\alpha\end{pmatrix}
\end{align}
\begin{figure*}[htb]
\epsfig{file=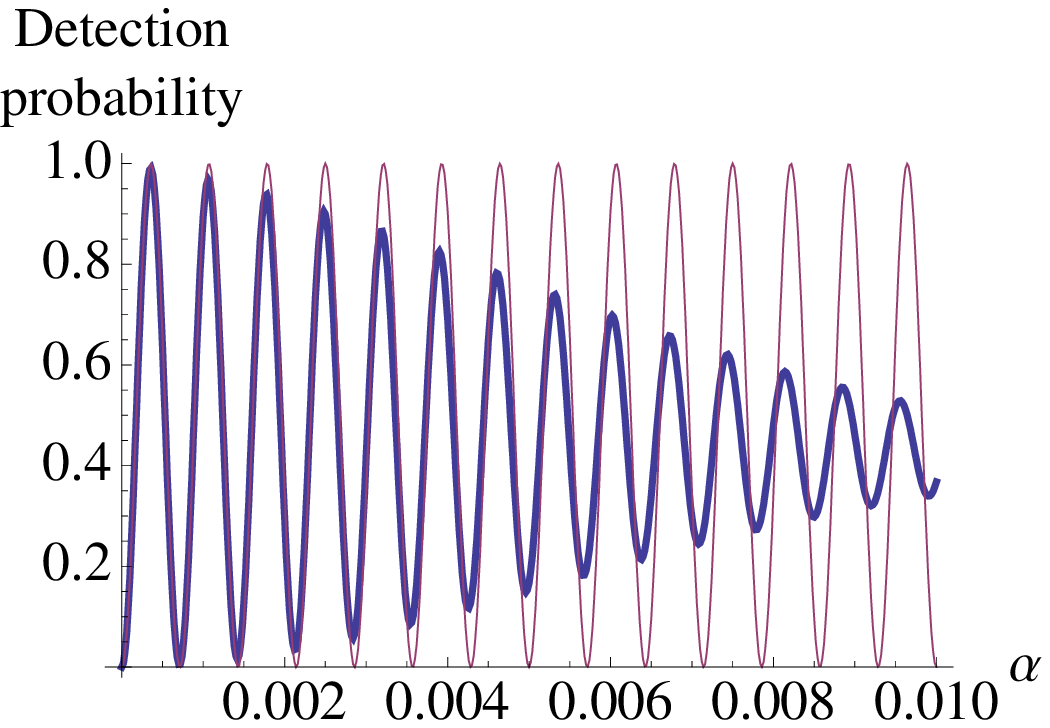,width=0.4\linewidth}\hspace{2cm}\epsfig{file=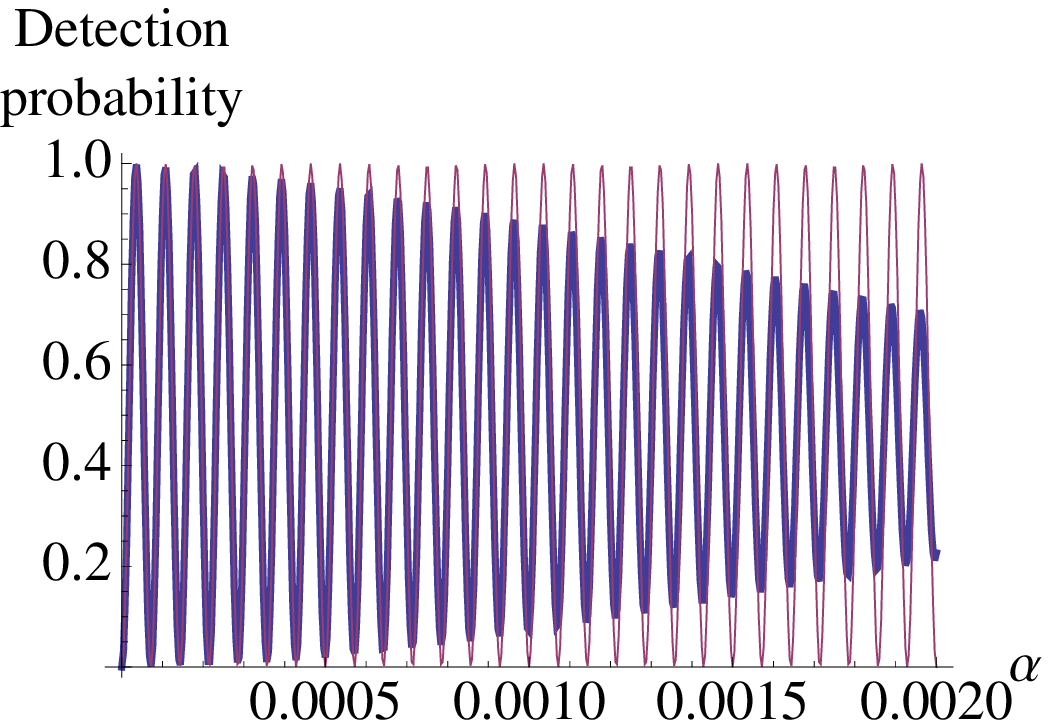,width=0.4\linewidth}
\caption[test]{\label{fig:output-1d}(Color online) Probability of observing an atom in the TT-RR output of the interferometer as a function of the tilt angle $\alpha$. The atoms propagate in a one-dimensional tight-binding potential period-two super-lattice with hopping energy $\frac12U$, on-site potential $\pm U$, and lattice spacing  $a=2\pi\hbar\sqrt{\frac{U}{m}}$. The thick blue line was calculated with the full expression (\ref{eq:ttfcrr}), and the thin violet is the small angle approximation (\ref{eq:oalpha-ex}). The initial momentum space width $\delta=\frac14\sqrt{\varepsilon}$ and the barrier width $w=\frac{1}{2\pi}\frac{a}{\sqrt\delta}$. The left and right panels show the interference patterns for $\varepsilon=10^{-3}$ and $\varepsilon=10^{-4}$, respectively.} \end{figure*}
\begin{figure*}[htb]
\epsfig{file=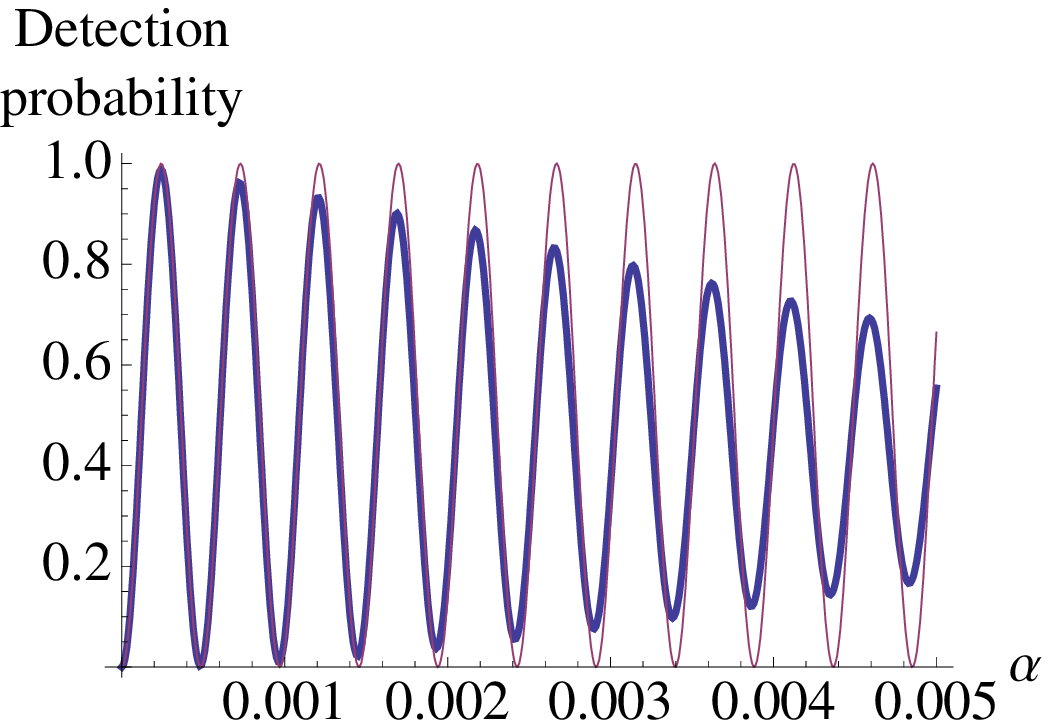,width=0.4\linewidth}\hspace{2cm}\epsfig{file=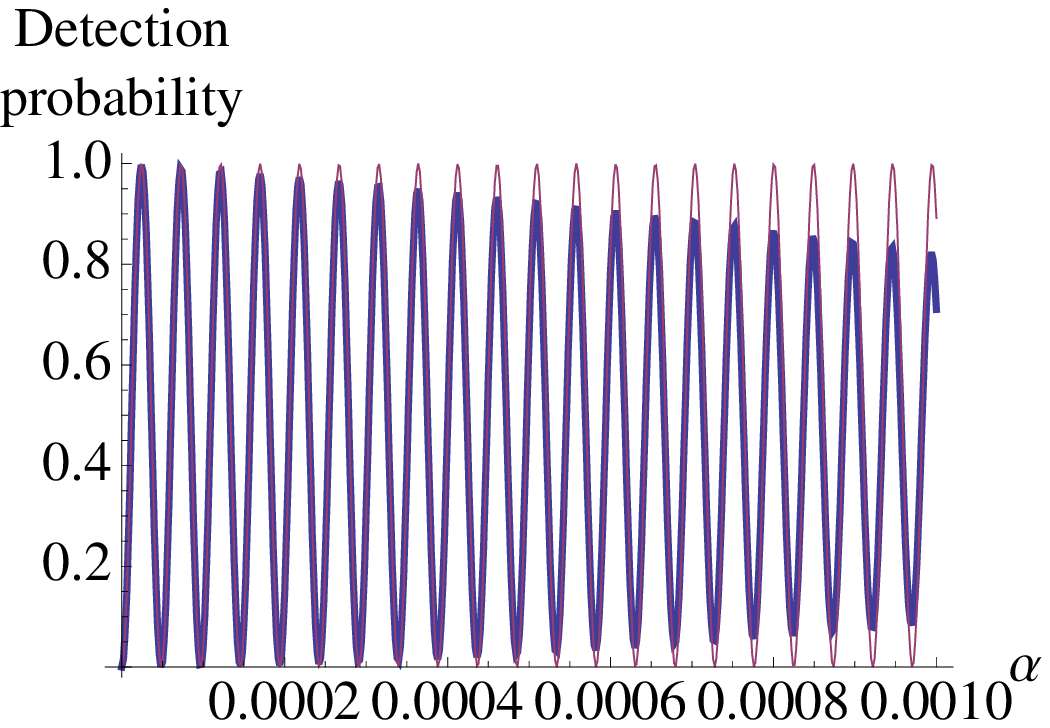,width=0.4\linewidth}
\caption{\label{fig:output-2d}(Color online) The interferometer output for propagation in a separable two-dimensional optical lattice that is a superposition of $x$ and $y$ potential super-lattices having the same properties as the one used in Fig.\ \ref{fig:output-1d}. The significance of the axes, the curves, the initial state, the width of the barrier, and the values of $\varepsilon$ are the same as in Fig.\ \ref{fig:output-1d}}
\end{figure*}
The state in the TT-RR output is $\ket{f}^{TT}+\ket{c}^{RR}$, so that the probability that an atom is detected there is $|t(\vp_c^{(T)})t(\vp_i)|^2+|r(\vp_c^{(R)})r(\vp_i)|^2+2\re{}^{TT}\bracket{f}{c}^{RR}$. The interference term takes the form
\begin{align}
{}^{TT}\bracket{f}{c}^{RR}&=t(\vp_c^{(T)})^*t(\vp_i)^*r(\vp_c^{(R)})r(\vp_i)\nonumber\\&
\times e^{\frac{2\pi i}{\varepsilon}\bigl(\int_{\vp_{i,x}}^{\vp_{f,x}^{(TT)}}\Ev(p_x,p_{i,y})dp_x -\int_{\vp_{i,x}^{(R)}}^{\vp_{c,x}^{(R)}}\Ev(p_x,p_{i,y}^{(R)})dp_x\bigr)}\nonumber\\&\textstyle\times \frac{e^{-\frac{i}{\hbar}(\vq_f^{(TT)}\cdot\vp_f^{(TT)}-\vq_i\cdot\vp_i)}e^{\frac{i}{\hbar}(\vq_c^{(R)}\cdot\vp_c^{(R)}-\vq_i\cdot\vp_i^{(R)})}}{\sqrt{\det(J_i)\det(J_c^{(R)})\det(\frac{1}{2}((C^{(TT)})^*+C^{(RR)}))}} \nonumber\\&
\times e^{-\frac{1}{4\delta^2}(\vp_f^{(TT)}\cdot(C^{(TT)})^*\cdot\vp_f^{(TT)}+\vp_c^{(RR)}\cdot C^{(RR)}\cdot\vp_c^{(RR)})} \nonumber\\&
\times e^{\frac{1}{4\delta^2}v\cdot((C^{(TT)})^*+C^{(RR)})^{-1}\cdot v}\label{eq:ttfcrr}
\end{align}
where $$v=\frac{2i\delta^2}{\hbar}(\vq_f^{(TT)}-\vq_c^{(R)})+(C^{(TT)})^*\cdot\vp_f^{(TT)}+C^{(RR)}\cdot\vp_c^{(RR)}$$
The phase of the interference term changes as a function of $\alpha$ on a scale of $\varepsilon$, while the amplitude and the frequency of phase oscillations change more slowly. These properties are visible in the interference patterns shown by thick blue lines in Figs.\ \ref{fig:output-1d} and \ref{fig:output-2d} for one- and two-dimensional optical lattices, respectively. The interference patterns were calculated for two values of $\varepsilon$, $10^{-3}$ and $10^{-4}$.

The results presented in this subsection were derived under the assumption that the T and R wave packets do not scatter before the recombination event. Since the trajectories of the wave packets in the interferometer arms cross the barrier line this requires temporal control of the barrier amplitude so that it vanishes at the intermediate crossing times of the T and R wave packets, and kept constant at the initial splitting time and final recombination times. If the barrier amplitude is not controlled temporally, further splitting would make the final state a superposition of six (rather than four) wave packets or more. Nevertheless the interference in the TT-RR and TR-RT channels persist with the modification that the state $\ket{f}^{TT}$ is multiplied by an additional transmission amplitude $t(\vp_x^{(T)})$ compared with Eqs.\ (\ref{eq:ftt}) and (\ref{eq:pftt}), where $\vp_x^{(T)}$ is the momentum of the T wave packet when it crosses the barrier, and similarly $\ket{c}^{RR}$ gains an additional factor of $t(\vp_x^{(R)})$ with respect to Eqs.\ (\ref{eq:crr}) and (\ref{eq:pcrr}). In the following we assume that the barrier potential is controlled appropriately and therefore omit these additional factors that reduce the amplitude of the interferometric oscillations in Eq.\ (\ref{eq:ttfcrr}), but otherwise do not affect the results.

\subsection{Interferometry for small angles}
The interference is effective when the interference term (\ref{eq:ttfcrr}) is large, and this requires that the wave packets $\ket{f}^{TT}$ and $\ket{c}^{RR}$ overlap in position as well as in momentum. For most tilt angles $\vq_f^{(TT)}\ne\vq_c^{(RR)}$ and $\vp_f^{(TT)}\ne\vp_c^{(RR)}$. However, if the beam splitter is aligned with the $y$ axis then by symmetry $p_{i,x}^{(R)}=-p_{i,x}$, $p_{i,y}^{(R)}=p_{i,y}$ and $v_y(-p_x,p_y)=v_y(p_x,p_y)$, so that after one Bloch period the two sub-wave packets meet at the beam splitter at $q_{c,x}^{(T)}=q_{c,x}^{(R)}=q_{i,x}$, $q_{f,y}^{(T)}=q_{f,y}^{(R)}=q_{i,y}+\frac1F\partial_y\bar\Ev(p_{i,y})$ where $\bar\Ev(p_{i,y})=\int_0^1 dp_xv_y(p_x,p_{i,y})$. There is therefore full overlap between the TT and the RR wave packets for $\alpha=0$, and high interferometric visibility for small enough $\alpha$.

Since $\alpha$ is small, it is possible to approximate the classical data in the interference term by its Taylor expansion. The $O(\alpha)$ approximation gives
\begin{align}\label{eq:oalpha}
&{}^{TT}\bracket{f}{c}^{RR}=\textstyle (t(\vp_i)^*r(\vp_i))^2(1+\alpha(h_1+h_2))\\
&\times e^{-\frac{2\pi i}{\varepsilon} \alpha\partial_\alpha p_{i,y}^{(R)}\partial\bar\Ev(p_{i,y})}\nonumber
\end{align}
where
\begin{align}
h_1&= \textstyle\partial_\alpha p_{c,x}^{(T)}\frac{\partial_x t(\vp_i)^*}{t(\vp_i)^*}+\partial_\alpha p_{c,x}^{(R)}
\frac{\partial_x r(\vp_i)}{r(\vp_i)}\nonumber\\
h_2&= \textstyle-\frac12\partial_\alpha(\det(J_i)\det(J_c^{(R)})\det(\frac{1}{2}((C^{(TT)})^*+C^{(RR)})) \nonumber
\end{align}

The second correction to the interference term $\alpha h_2$ is $O(\alpha)$ and therefore negligible for small $\alpha$; similarly, $\alpha h_1=O(\frac{w}{a}\alpha)$ is also negligible throughout the region of validity of Eq.\ (\ref{eq:oalpha}), where, therefore,
\begin{equation}\label{eq:oalpha-ex}
\bracket{f}{c}^{RR}=\textstyle (t^*r)^2
e^{\frac{4\pi i}{\varepsilon}\partial\bar\Ev(p_{i,y})\alpha}
\end{equation}
using the shorthand $t=t(\vp_i)$, $r=r(\vp_i)$.
It follows that for small enough angles the probability of observing the atom in the TT-RR output oscillates sinusoidally as a function of $\alpha$ on a fast scale of $O(\varepsilon)$ and large visibility. The thin violet lines in Figs.\ \ref{fig:output-1d} and \ref{fig:output-2d} show the interference pattern with the approximation Eq.\ (\ref{eq:oalpha-ex}) that becomes increasingly accurate for smaller $\alpha$.
If the band dispersion is known, these oscillations can be used to measure the force $F$ with high precision, as discussed further below. 

The domain of validity of Eqs.\ (\ref{eq:oalpha}) and (\ref{eq:oalpha-ex}) is limited by the size of the neglected $O(\alpha^2)$ terms. These terms arise in the combinations $\frac{\alpha^2}{\delta^2}$, $\frac{\alpha^2}{\varepsilon}$, and $\frac{\alpha^2\delta^2}{\varepsilon^2}$ in Eq.\ (\ref{eq:ttfcrr}). In particular, the first and third combinations determine the scale of decay of visibility as a result of momentum mismatch and position mismatch between the TT and RR wave packets, respectively. It follows that the domain of validity is
\begin{equation}\textstyle
\alpha\ll\min(\delta,\varepsilon^{\frac12},\frac{\varepsilon}{\delta})
\end{equation}
Evidently, the largest range is obtained when $\delta\sim\varepsilon^{\frac12}$, that is when the momentum and position uncertainties are balanced, so that the domain of validity is $\alpha\ll \varepsilon^{\frac12}$. In this case, since the oscillation period is $O(\varepsilon)$, the number of equal period oscillations up to a fixed tolerance is $O(\varepsilon^{-\frac12})$. When $\alpha$ is comparable with $\varepsilon^{\frac12}$ the visibility of the interference pattern decreases and the oscillation period changes, as shown in Fig.\ \ \ref{fig:output-1d} and \ref{fig:output-2d}, that also demonstrate that the number of high-visibility constant-period oscillations increases when $\varepsilon$ becomes smaller.

\section{Conclusions}\label{sec:cc}
An experiment of angle interferometry would detect the fraction of atoms scattered in the TT-RR arm for variable tilt angle of the tunnel barrier, keeping the rest of the parameters fixed. Writing Eq.\ (\ref{eq:oalpha-ex}) in physical units, this fraction would 
be approximately equal to the probability
\begin{equation}\label{eq:oalpha-phys}
\bracket{f}{c}^{RR}=\textstyle |t|^4+|r|^4+2\re((t^*r)^2
e^{2 i\frac{\alpha}{\hbar F}\partial\bar\Ev(p_{i,y})})
\end{equation}
With the optimal choice of initial uncertainty this expression is valid for values of $\alpha$ significantly smaller than $\varepsilon^{-\frac12}$.
A time of flight measurement can yield the squared magnitude of the full momentum-space wave function, the TT-RR part of which is given by Eqs.\ (\ref{eq:pftt}) and (\ref{eq:pcrr}).

The interferometer period can be further simplified when the optical lattice is separable, a case that includes one-dimensional lattices, obtaining
\begin{equation}\label{eq:oalpha-sep}
\bracket{f}{c}^{RR}=\textstyle |t|^4+|r|^4+2\re((t^*r)^2
e^{2 i\frac{v_{i,y}}{\hbar F} \alpha})
\end{equation}
where $v_{i,y}$ is the constant velocity in the direction perpendicular to the force. 

If gravity is used as the external force acting on atoms with mass $m$ in lattice with spacing $a$ then
\begin{equation}
\varepsilon=\frac{m^2a^3g}{u(\pi\hbar)^2}
\end{equation}
where $g$ is the acceleration of gravity, and $u$ is the band width in units of the recoil energy $\frac{(\pi\hbar)^2}{a^2m}$. It follows that the interferometric condition $\varepsilon\ll1$ is experimentally accessible. For example, for Sodium atoms in a $\frac12589\,\text{nm}$ lattice $\varepsilon=\frac{1}{u}3.3\times10^{-2}$; a lattice depth of $0.25$ recoil energy, that can be set up with sub-mW lasers \cite{lat}, gives $u\approx0.4$, and $\varepsilon=8.4\times10^{-2}$. Spatial localization of $3\,\mu$m and barrier width of $1\,\mu$m can be chosen to satisfy the basic inequalities Eq.\ (\ref{eq:ie}).

A natural application of the interferometer is force measurement, that can reach high precision thanks to the high sensitivity of the interference pattern, and for this application a one-dimensional lattice suffices, and the force can be deduced from the interferometer output using Eq.\ (\ref{eq:oalpha-phys}). The statistical error in a single measurement is of the order of the inverse square root of the number of atoms, that can be $10^{-3}$ in typical experiments. If the measurement is repeated for one hundred values of $\alpha$ the statistical error is reduced by another factor of 10. The induced measurement error in the force is reduced by the number of visible interferometric oscillations that is of order $\sqrt{\varepsilon}$, giving an error estimate of $10^{-5}$. The interferometer measures the combination $\frac{F}{v_{i,y}}$ rather than the force itself, so that high-precision measurement of the transverse speed is needed to reach this value of force measurement sensitivity. This sensitivity is not as good as that of existing interferometers that rely on the classical Bloch oscillation phase, but the dynamical phase interferometer has the advantage of requiring a \emph{single} Bloch oscillation, facilitating much faster measurements.

The interferometer can also be used to characterize the dispersion of two-dimensional lattices with a given external force. This application can become especially interesting if the lattice breaks space-reflection symmetry so that the interferometer can measure a non-trivial Berry phase \cite{zak89}. This question, however, requires a more accurate analysis that is beyond the scope of the present paper.

\emph{Acknowledgments}: We benefited from informative and helpful discussions with Nir Davidson and Nadav Katz. This research was supported by the German-Israeli foundation under grant 980-184.14/2007.

\end{document}